\documentclass{article}
\usepackage{fullpage}
\usepackage{amsmath}
\usepackage{amssymb}
\usepackage{amsthm}
\usepackage{algorithm}
\usepackage[noend]{algpseudocode}
\usepackage{graphicx}
\graphicspath{{img/}}
\usepackage{parskip}
\usepackage{subfigure}
\usepackage{color}
\usepackage{array}
\newcolumntype{L}[1]{>{\raggedright\let\newline\\\arraybackslash\hspace{0pt}}m{#1}}
\title{A Computational Study of Feasible Repackings\\
	in the FCC Incentive Auctions}
\author{Michael Kearns 
	and Lili Dworkin
	\footnote{Research conducted on behalf of AT\&T. All experiments, analyses, exposition, and opinions are exclusively the work of the authors.
		Contact author M. Kearns may be reached at \tt{mkearns@cis.upenn.edu}}\\
	Computer and Information Science\\
	University of Pennsylvania
}
\date{}

\newcommand{\ld}[1]{\textcolor{blue}{[LD: #1]}}
\newcommand{\mk}[1]{\textcolor{red}{[MK: #1]}}

\begin{document}
\maketitle

\begin{abstract}
We report the results of a computational study of repacking in the FCC Incentive Auctions.
Our interest lies in the structure and constraints of the solution space of feasible
repackings. Our analyses are ``mechanism-free'', in the sense that they identify constraints that
must hold regardless of the reverse auction mechanism chosen or the prices offered for broadcaster
clearing. We examine topics such as the amount of spectrum that can be cleared nationwide, 
the geographic distribution of broadcaster clearings required to reach a clearing target, 
and the likelihood of reaching clearing targets under various models for broadcaster participation.
Our study uses
FCC interference data and a satisfiability-checking approach, and elucidates both
the unavoidable mathematical constraints on solutions imposed by interference, as well
as additional constraints imposed by assumptions on the participation
decisions of broadcasters. 
\end{abstract}

\section{Introduction}

We report the results of an extensive computational study of the broadcaster repacking problem,
which plays a central role in the FCC's proposed Incentive Auctions. The Incentive Auctions are designed to 
repurpose broadcast television spectrum for wireless broadband use, and the reverse auction portion necessitates
that broadcasters not cleared by the process be repacked onto lower frequencies. Co-channel and adjacent
channel interference constraints, as well as those presented by international borders and other issues,
make repacking a computationally challenging task.
\footnote{Throughout the paper, we shall assume the reader is familiar with the Incentive Auctions,
and in particular the nature of the repacking problem in the reverse auction and the formulation of its feasibility as an
instance of Boolean formula satisfiability.}

While the FCC is currently considering a particular descending clock mechanism for the reverse auction, as well
as other potential mechanisms,
the problem of
determining which subsets of broadcasters can be feasibly repacked at a given nationwide clearing target 
\footnote{Our focus on nationwide clearing targets is done for simplicity of exposition;
we offer no opinion on whether a nationwide or more variable approach is appropriate for any resulting wireless band plan.}
is one that must be addressed by {\em any\/} mechanism. The structure and constraints on this solution
space are thus ``mechanism-free'' and of fundamental interest, and they are the subject of our study.
Examples of the types of questions we pose and investigate here include:
\begin{itemize}
\item {\em For a given nationwide clearing target, what is the minimum number of broadcasters that must
	be cleared in order for there to exist a feasible repacking solution for the remaining broadcasters,
	and how does this number increase with higher clearing targets?}
\item {\em For a given nationwide clearing target, how many broadcasters must be cleared from each
	local market in order for there to exist a feasible repacking solution, and how variable is
	this number from solution to solution?}
\item {\em What is the cost of border constraints in terms of the number of broadcasters cleared and
	its geographic distribution?}
\item {\em How diverse is the space of possible feasible solutions at a given clearing target?}
\item {\em How do the answers to such questions change under varying assumptions on the
	participation decisions of different classes of broadcasters (such as network affiliates)?}
\end{itemize}

To investigate these questions and related ones, we have implemented a computational framework
in which the feasibility of repacking is formulated as an instance of the satisfiability of a
Boolean formula that captures the nationwide interference constraints provided by FCC data, and the existence of solutions is determined using the open-source software package PicoSAT. Our
experimental platform is thus very similar to that used in recent intensive FCC studies, but our purpose
is complementary: whereas the FCC studies have naturally focused on greatly improving the computational
efficiency of determining repacking feasibility for the real-time needs of a live auction,
our focus here is on elucidating the fine-grained structure of the solution space, and the constraints
on it; computational efficiency is only a secondary concern.

We wish to emphasize from the outset that all of our findings are the results of
logical and numerical computations performed using publicly released FCC data, and in some cases,
models and assumptions for the probabilities of participation by broadcasters and groups of broadcasters.
Consequently {\em all results and conclusions reported here should be viewed as contingent on the
quality and accuracy of that data, and the validity of those assumptions.}
We make no claims about the generalization of our results to other data sets or assumptions, and for
this reason we encourage the reader to focus more on the qualitative and relative findings
(such as the rate at which broadcaster clearings must increase with clearing target, or the geographic
locations of broadcaster clearings and their relative distribution), as opposed to the precise
numerical values we do indeed provide. The qualitative and relative findings are more likely to survive plausible changes to the
data (such as FCC revisions to interference and border constraints) and assumptions (such as alternate
broadcaster participation models); the precise numerical findings are less likely to.

\subsection{Form of Analyses}

Our analyses fall into two categories that can be
thought of as what ``must'' happen in the repacking solution space, and what ``might'' happen.
In the ``must'' category, we assume that in principle every broadcaster is willing to
relinquish their license (at some price), and investigate solution space structure such as the
smallest number of broadcaster clearings required nationwide, or on a market-by-market basis, at a given
clearing target. One can view these as thought experiments into the most clearing-efficient solutions available, acknowledging 
that they do not account for practical considerations on participation. We refer to these as
``must'' or absolute analyses because they investigate purely mathematical and structural constraints on the solution
space; there is simply no getting around these constraints, regardless of the mechanism or the
prices it offers.
\footnote{Note that all of our
analyses deliberately avoid any assumptions about 
clearing prices or any other mechanism-specific details.}

These analyses are not only mechanism-free, they are also ``model-free'' regarding participation
decisions by individual broadcasters, since we explicitly assume any broadcaster is willing to clear.
While this is valuable in identifying the absolute constraints on
repacking solutions, it is unrealistic in the sense that we may have prior knowledge or beliefs
about the likelihood of participation by certain broadcasters or classes of broadcasters, such as network
affiliates. In other words, rather than only identifying what ``must'' happen purely from solution
space constraints, we also interested in what ``might'' happen under various participation models.
These models are not meant to reflect predictions or judgments about participation, but rather to examine hypothetical 
outcomes under varying participation assumptions.

The participation models we examine take the form of a joint probability distribution over participation
decisions by broadcasters; more formally, we represent each broadcaster's availability for
clearing (again, at some unspecified price) as a binary random variable. The joint distribution
of these variables for all broadcasters nationwide can then be designed to capture prior beliefs
about correlations in participation across groups of broadcasters. 

To give perhaps the simplest example, suppose one
believed that roughly a fraction $\alpha$ of all broadcasters will choose not to participate, and that each broadcaster
will make an independent, random participation decision. This (perhaps naive) assumption could be modeled
by the standard binomial distribution 
$$P(\vec{x}) = \prod_{i=1}^n \alpha^{x_i} (1-\alpha)^{1-x_i}$$
which simply corresponds to a distribution in which each broadcaster $i$ flips a (biased) coin with probability
$\alpha$ of coming up heads ($x_i=1$, representing non-participation and thus forcing repacking) 
and $1-\alpha$ of coming up tails ($x_i=0$, representing participation or availability for clearing).
Already for this most basic model of participation, one can ask a computationally and analytically
non-trivial question of interest: as a function of $\alpha$, how much spectrum might one expect to clear
nationwide? Computationally, this can be answered by drawing samples $\vec{x}$ repeatedly
from the distribution $P$, and determining via satisfiability testing whether the set of broadcasters $i$ for which
$x_i = 1$ can be feasibly repacked at various clearing targets. Thus such analyses are methodologically
very similar to the ``must'' analyses above, but have a crucial conceptual difference: while the ``must''
analyses identify unavoidable mathematical constraints on repacking, here we have a model $P$
for participation decisions, and any conclusions are of course dependent on the model chosen.

In principle our framework can accommodate any joint distribution 
$P(\vec{x})$, allowing us to explore scenarios in which the participation decisions of various groups
of broadcasters are correlated in varying ways and strengths. For example, one might posit that network
affiliate (e.g. ABC, NBC, CBS) broadcasters might make decisions that are more
likely to be in agreement with each other than the overall population of broadcasters. Such beliefs can be 
modeled by the introduction of hidden variables in the form of a
distribution that correlates affiliates of a specific network or affiliates in general in a
hierarchical fashion.
Similarly, we can examine joint participation distributions
that correlate participation behavior with broadcaster revenue or other economic,
geographic or demographic properties of broadcasters. For instance, one can posit distributions
in which the most profitable broadcasters are less likely to participate, or in which the
participation of a coalition of broadcasters with a shared corporate parent are correlated.
For any proposed model
$P(\vec{x})$ (which we view as the input to such analyses, with all subsequent findings
being conditioned on the plausibility and realism of the input model), our broad methodology
remains the same: for various clearing targets, we estimate the probability of being able to
successfully clear that target (repack the non-participants), and perform a variety of related analyses.

\section{Interference Data and Computational Framework}

To begin, we formally define the repacking problem, and describe our data and experimental framework. We run our experiments on a dataset obtained by filtering the US baseline provided by the FCC on July 22\textsuperscript{nd} 2013. 
\footnote{
Recently (June 2 2014), the FCC released an updated dataset
that replaces the use of proxy channels 
with the specific pairwise channel assignments for
determining potential interference constraints, along with an analysis 
showing this should have a minimal impact on feasibility. While our analysis does
not use this updated release, we similarly expect it would have minimal impact on
our analyses. 
(FCC Public Notice DA 14-677)}
We only consider the 1672 broadcasters currently assigned to UHF channels (14-51). We ignore low and high VHF stations because they are unlikely to be repacked onto the UHF band. The goal of the repacking process is to assign broadcasters to a subset of the 38 channels in the UHF band, and therefore clear a large contiguous block of spectrum. Because each channel uses six MHz of spectrum, reaching a clearing target of $M$ MHz is equivalent to clearing $\lfloor M/6 \rfloor$ channels. This leaves $38 - \lfloor M/6 \rfloor$ channels available for repacking. \footnote{If the clearing target is greater than 84 MHz, we in fact need to clear $\lfloor M/6 \rfloor + 1$ channels, because channel 37 is reserved for public safety broadcast. Thus, there will only $37 - \lfloor M/6 \rfloor$ channels available for repacking.} A solution $A$ to the repacking problem consists of an assignment of broadcasters to the remaining $c$ channels. Let $A(i) = k$ denote that the solution assigns broadcaster $i$ to channel $k$. A feasible solution must obey a number of \emph{interference constraints}. The FCC has used specialized software (OET-69) to decide, for every pair of stations, whether the pair can be assigned to the same or adjacent channels. The results were publicly released on July 22\textsuperscript{nd} 2013 in the form of binary matrices. There are three types of interference constraints. A \emph{co-channel} constraint indicates that two broadcasters $i, j$ cannot be assigned to the same channel in an assignment, or equivalently $A(i) \neq A(j)$. 
An \emph{adj-channel-up} constraint indicates that broadcaster $i$ cannot be placed on the channel above another broadcaster $j$, i.e. $A(i) \neq A(j) + 1$, and an \emph{adj-channel-down} constraint indicates that broadcaster $i$ cannot be placed on the channel below another broadcaster $j$, i.e. $A(i) \neq A(j) - 1$. 

A feasible solution must also obey a set of \emph{domain constraints}, each of which enforces that a broadcaster $i$ cannot be assigned to a particular channel $k$, i.e. $A(i) \neq k$. The majority of domain constraints are a result of international treaties with Canada and Mexico, which forbid US stations from being assigned to channels on which they would interfere with stations across the border. The FCC has also released a description of these constraints in the form of a binary matrix. Because these current domain constraints may or may not restrain the ultimate auction, we have run our experiments both including and excluding these constraints. The reality will therefore lie somewhere between the two results. 
We note that all of our reported results forbid any assignment or repacking onto the public safety broadcast channel 37, even when border and other domain constraints have been dropped. However, when domain constraints are excluded, other non-border domain constraints,such as those involving T Band public safety channels, are excluded as well.

Given a clearing target $M$ and a set $S$ of broadcasters that must be repacked, our goal is to find a feasible channel assignment. We first encode the constraints described above as Boolean formulae. In particular, we write formulae to ensure that:
\begin{enumerate}
\item Each broadcaster in $S$ is repacked onto exactly one channel.
\item $38 - \lfloor M/6 \rfloor$ channels are not occupied by any broadcaster.
\item All interference and domain constraints are satisfied. 
\end{enumerate}
There are some optional parameters as well; i.e. we can specify a maximum number $b$ of broadcasters to clear nationwide, a maximum number $b'$ of broadcasters to clear from a particular DMA, and a maximum number $d$ of DMAs in which clearing is allowed to occur. Each of these additional constraints can also be encoded as Boolean formulae. The complete encoding is provided in the Technical Appendix in Section \ref{sec:cnf}. We then give these formulae as input to PicoSAT, an open-source SAT solver initially
developed in 2007 by Armin Biere at Johannes Kepler University in Austria. If a feasible solution exists, PicoSAT will typically find a corresponding channel assignment in 30 seconds or less. If the constraints are unsatisfiable, however, PicoSAT may or may not quickly find proof of infeasibility. In our experiments, we assume infeasibility after 60 seconds have elapsed. 

Although we have run the reported experiments on a wide range of clearing targets,
and both with and without domain constraints,
in the interests of brevity
we sometimes choose to report results only for an 84 MHz clearing target without domain constraints, since this
target has been frequently used as an example in other writings about the Incentive Auctions. We have no opinion on the optimal or appropriate clearing target for the auction.

\section{Absolute Constraints on the Solution Space}

\subsection{Clearing Requirements at the National Level}

We begin our results with perhaps the most basic of ``must'' analyses: if we assume that every
broadcaster is willing to relinquish their license at some unspecified price, what is the absolute {\em minimum\/}
number of broadcasters that must be cleared in order to meet a given nationwide target?

\begin{figure}[]
\centering
\subfigure{\includegraphics[width=0.65\textwidth]{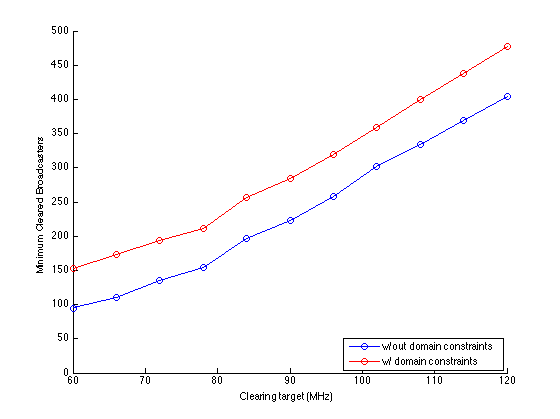}}
\caption{\it Minimum number of broadcasters cleared as a function of nationwide clearing target, in order
	to permit a feasible repacking; both including and ignoring domain constraints. 
}
\label{fig:absmin}
\end{figure}

The answer is provided in Figure~\ref{fig:absmin}. To generate this figure, we encoded Boolean
formulae that specified not only the clearing target, but also a maximum number $x$ of broadcasters
allowed to be cleared in the solution. By gradually reducing $x$ for a given clearing target until the
point that the formula is determined to be infeasible, we identify the smallest number of clearings required.
In the figure, we show plots both incorporating and ignoring the domain constraints.

As can be seen in Figure~\ref{fig:absmin},
the minimum number of clearings required increases approximately linearly, and reasonably gradually, with the nationwide target;
even this fact is not self-evident from first principles. Specific values of interest
can be read directly from the plot, such as slightly less than 200 broadcaster clearings being required
for clearing an 84 MHz target ignoring border constraints. The gap between the two curves reveals that
the cost of the currently released domain constraints is approximately 50 additional broadcaster clearings,
regardless of the nationwide target. If one believes the current domain constraints to be overly
conservative, this cost would presumably be lower.

We again emphasize that in this analysis, {\em all\/} broadcasters are considered as candidates for clearing --- in effect,
we are letting the satisfiability solver choose the particular set that is cleared in service of reaching the given 
clearing target. At lower clearing targets, there are many possible solutions with many possible sets of cleared broadcasters;
at higher targets, these choices become more constrained. We shall comment further on the diversity of the solution space below.

\subsection{Clearing Requirements at the DMA Level}

Figure~\ref{fig:absmin} identifies the most basic and nationwide constraints on clearing a given
amount of spectrum.
We next report on absolute clearing constraints on a finer geographic scale, namely at the
level of individual markets or DMAs. More specifically, in order to achieve a given nationwide target 
efficiently (in the sense of Figure~\ref{fig:absmin}),
and again assuming every broadcaster is a candidate for clearing, how many broadcasters must be cleared
in each market?

A meaningful answer to this question requires that we make it a bit more precise. First, since we are
interested primarily in solutions that are efficient in the number of clearings nationwide, we encode the
constraint that this number be not too much larger (say, 5\%) than the values dictated by Figure \ref{fig:absmin}
\footnote{The reason for allowing a slight buffer over the absolute minimum number of clearings is discussed below.};
otherwise, we permit the trivial solution in which all broadcasters are cleared and essentially any target can be
reached. The more subtle issue is how to constrain clearings on a per-DMA basis.
This could be done by adding constraints that specify 
a maximum number $x$ of broadcasters allowed to be cleared in a particular DMA, and gradually reduce $x$ until 
infeasibility occurs. Such an approach would determine the minimum number of broadcasters that 
must be cleared in this DMA to achieve feasibility; we could then repeat for each DMA. 
The problem with this approach is that there is no guarantee that all the individual per-DMA minima found can be 
achieved {\em simultaneously\/} in a single feasible solution --- indeed, this is certainly not the case. 
For instance, clearing the minimum number of broadcasters in one DMA will 
frequently necessitate that extra broadcasters are cleared in neighboring DMAs, a consequence of
so-called ``daisy-chain'' or constraint propagation effects. (We shall demonstrate these 
effects in a precise sense shortly.) 

Thus at the DMA level, it is more informative to examine ``typical'' nationwide feasible solutions and 
then measure the average number of broadcaster clearings occur in each DMA. More precisely, for a given
nationwide clearing target, and fixing the total number of broadcasters cleared at (approximately) the minimum possible
as given in Figure~\ref{fig:absmin},
there will in general be {\em many\/} feasible repacking solutions that vary in the
geographic distribution of cleared broadcasters. Because the algorithm used by PicoSAT is
randomized, we can run it multiple times in order to sample this solution space.
\footnote{PicoSAT employs a greedy search heuristic in which 
a satisfying assignment of the input formula, if one exists, is gradually constructed
from an initial assignment. Since this process periodically involves setting variables
randomly when there is a choice for their values,
by running the algorithm repeatedly on the same formula, we can sample the space of feasible solutions.
While we cannot claim that the samples are distributed {\em uniformly\/}, the process does provide a good sense of the variability across feasible solutions, which is the main goal of our analysis here. Shortly we shall quantify the size and diversity of the solution space. Although producing uniform samples for a given boolean formula is known to be computationally intractable, it is possible to samply near-uniformly from the set of feasible solutions. Gomes et. al. provide such an algorithm in ``Near-Uniform Sampling of Combinatorial Spaces Using XOR Constraints,'' Proceedings of Neural Information Processing Systems, 2006. Incorporation of this approach is a possibility for future work.
}

\begin{table}
\resizebox{\textwidth}{!}{
\begin{tabular}{l L{2in} l l l l l}
Rank & DMA & Size & Avg \# Cleared & Std Dev & Observed Min & Absolute Min\\ \hline
1  & New York, NY                                  & 23 & 10.177  & 1.229   & 8 & 1 \\
2  & Philadelphia, PA                              & 23 & 8.31    & 1.2804  & 5 & 0 \\
3  & Chicago, IL                                   & 21 & 8.1     & 1.3276  & 5 & 1 \\
4  & San Francisco - Oakland - \newline San Jose, CA            & 23 & 7.8767  & 1.2301  & 2 \\
5  & Boston, MA                                    & 17 & 7.15    & 1.2051  & 3 & 0 \\
6  & Los Angeles, CA                               & 27 & 6.8733  & 0.98005 & 5 & 2 \\
7  & Orlando - Daytona Beach - Melbourne, FL           & 21 & 6.1967  & 1.2067  & 3 & 2 \\
8  & Miami - Ft. Lauderdale, FL                    & 18 & 5.7633  & 1.4402  & 2 & 0 \\
9  & Raleigh-Durham, NC                            & 17 & 5.44    & 0.93236 & 3 & 0 \\
10 & Milwaukee, WI                                 & 12 & 5.23    & 1.5137  & 1 & 0 \\
11 & Greensboro - High Point - Winston Salem, NC       & 9  & 4.8333  & 1.0144  & 2 & 0 \\
12 & Tampa - St Petersburg - Sarasota, FL              & 15 & 4.8     & 1.1154  & 2 & 0 \\
13 & Washington, DC                                & 17 & 4.69    & 1.1798  & 2 & 0 \\
14 & Charlotte, NC                                 & 15 & 4.6133  & 1.0036  & 2 & 0 \\
15 & Dayton, OH                                    & 7  & 4.1967  & 0.80839 & 2 & 0 \\
16 & Indianapolis, IN                              & 16 & 4.0033  & 0.96943 & 2 & 0 \\
17 & Baltimore, MD                                 & 7  & 3.6933  & 0.8092  & 1 & 0 \\
18 & Cleveland - Akron, OH                           & 15 & 3.6033  & 1.0277  & 1 & 0 \\
19 & Atlanta, GA                                   & 15 & 3.5833  & 0.91241 & 1 & 0 \\
20 & Columbus, OH                                  & 14 & 3.38    & 0.96888 & 1 & 0 \\
21 & West Palm Beach - \newline Ft. Pierce, FL                & 11 & 3.2033  & 1.1775  & 1 & 1 \\
22 & Pittsburgh, PA                                & 19 & 2.84    & 1.0762  & 0 & 0 \\
23 & Cincinnati, OH                                & 9  & 2.6433  & 0.85551 & 0 & 0 \\
24 & Denver, CO                                    & 16 & 2.6067  & 1.1874  & 0 & 0 \\
25 & Greenville - Spartanburg, SC - Asheville, NC      & 10 & 2.6033  & 0.87285 & 0 & 0 
\end{tabular}}
\caption{\it Top 20 DMAs when ranked by the average number of broadcasters that must be cleared in the DMA
	in order to reach a feasible solution for a nationwide 84 MHz clearing target, and ignoring domain constraints. 
}
\label{tab:dma}
\end{table}

In Table \ref{tab:dma}, we show sample results for exactly such an experiment. The data underlying this table
consisted of 300 sampled feasible repacking solutions at an 84 MHz clearing target and with the
total number of broadcaster clearings constrained to be at most 206 (10 larger than the minimum possible),
and ignoring domain constraints for the moment.
From these sampled solutions, we compute the average number
of broadcaster clearings within each of the 210 DMAs, as well as the standard deviations of this quantity.
We also report the minimum number of clearings observed in the sample for each DMA. Together these metrics
provide us with a geographically granular view of the space of feasible solutions.

Table \ref{tab:dma} shows the top DMAs when sorting by the average number of broadcaster clearings.
Unsurprisingly, at the top of the list we find large and ``congested'' markets like New York City, Philadelphia, and Chicago. 
But some less obvious challenge areas appear as well, such as large swaths of South Florida, North Carolina, and Ohio.
It is also noteworthy that in the highest ranked DMAs, the standard deviations are small compared to the averages themselves. 
For instance, in New York City, the average solution required about 10 broadcasters to be cleared, and the standard deviation is only about 1. 
Of the 300 sampled solutions, the minimum number of clearings observed in this DMA was 8.
This suggests that, at least statistically speaking (and with respect to the distribution over solutions generated by PicoSAT),
that there is very little flexibility in the amount of clearing required in these largest markets, especially when considered
simultaneously and in the aggregate.
In the lower ranked markets, the standard deviation is much closer to the mean, which suggests that there is considerably more flexibility in these markets.

\begin{figure}[]
\centering
\includegraphics[width=0.75\textwidth]{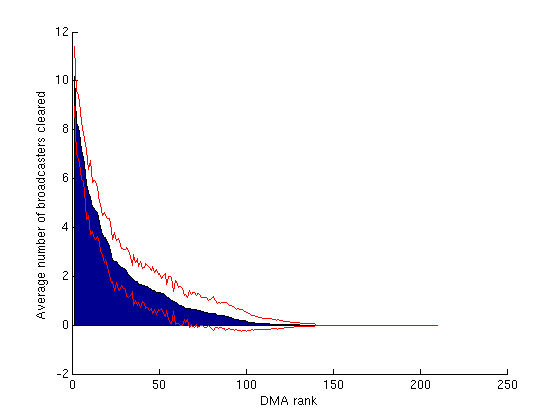}
\caption{\it Average number of broadcaster clearings per DMA at 84 MHz, sorted by the same quantity.
The upper and lower red curves show one standard deviation above and below the means, respectively.}
\label{fig:dmabar}
\end{figure}

\begin{figure}[]
\centering
\subfigure[Without domain constraints]{\includegraphics[width=1.0\textwidth]{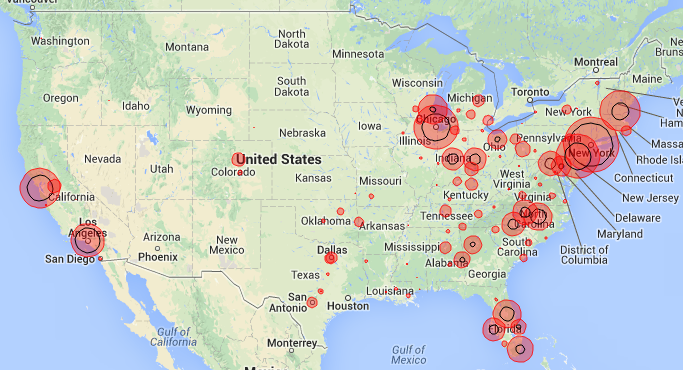}} \\
\subfigure[With domain constraints]{\includegraphics[width=1.0\textwidth]{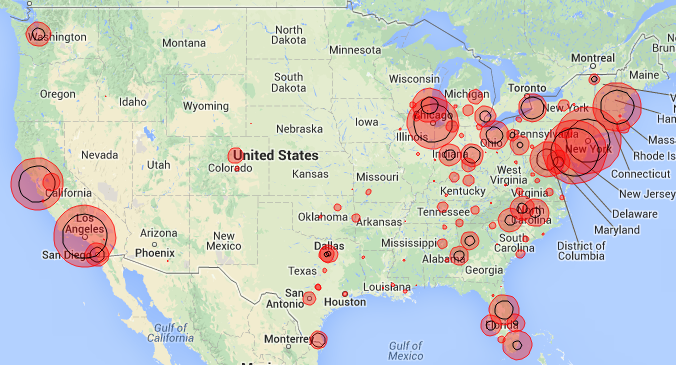}} \\
\caption{\it A map illustrating the average number of clearings per DMA, with and without domain constraints. 
	Each DMA is labeled by a red circle whose area is proportional to the average number of clearings required in the sample of 300 feasible solutions.
	The inner black circle area is proportional to the minimum number of clearings observed in the sample for the DMA.
}
\label{fig:map}
\end{figure}

Two different visualizations of the distribution of clearings required across all DMAs are provided in
Figures ~\ref{fig:dmabar} and \ref{fig:map}. 
Figure~\ref{fig:dmabar} visualizes the sorted average
number of clearings required per DMA, overlaid with the standard deviations.
Figure~\ref{fig:map} projects the same information onto a geographic map, both with and without domain constraints.

For comparison, we also used the alternate Boolean formulae approach discussed above to find the absolute 
minimum number of broadcasters that must be cleared in each DMA {\em in isolation\/}, while still requiring a near-optimal amount 
of clearing nationwide. Recall that these minima can certainly not be achieved simultaneously in a single 
solution. Rather, for each DMA, a solution does exist in which the minimum for that particular DMA is achieved. 
Interestingly, these absolute minimum single-DMA values are 0 for almost all DMAs, including many of the largest,
meaning
we can essentially avoid clearing {\em any\/} broadcasters in any particular DMA if we are willing to pay a higher price in neighboring markets.

\subsection{Daisy-Chain Effects and DMA Correlations}

Above we have argued that the absolute minimum number of clearings required in a specific DMA, or even
the minimum number observed in our sample for that DMA, may be misleading quantities for understanding
what can be plausibly achieved, and that the sample averages are more appropriate guides to expectations.
As we have intimated, this is due to what are sometimes referred to as ``daisy-chain'' effects, more
technically described as propagation over the network of network of interference constraints. The fundamental
issue is that if we clear less in one market, we may have to clear more in surrounding markets. Here we
provide some quantitative analysis of such effects. 

Since each of our 300 sampled solutions specifies how many clearings were required in each DMA to reach
the nationwide clearing target, we can examine which pairs of DMAs have the strongest correlations --- positive
or negative --- in terms of number of clearings. If daisy-chain effects are present and strong, we would
expect the most negative correlations to be large in magnitude, and between geographically proximate DMAs.

This expectation is strongly borne out by Table~\ref{tab:corrpos}, where we list the topmost negatively correlated
DMA pairs.
\footnote{We restrict attention only to DMAs with a significant amount of clearing --- namely,
at least 2 broadcasters cleared on average in the sample. Including DMAs with little or no clearing
leads to many spurious correlations, especially positive ones. All correlations reported are significant
at the $P \leq 0.01$ level.}
These pairs are all geographically adjacent DMAs, and also generally in highly congested regions requiring
large amounts of clearing. Thus while it is clear that in any individual DMA we may observe (or be able to
enforce) a small amount of clearing, in congested areas this only ``pushes the problem elsewhere'' in a quantifiable sense. 

\begin{table}
\resizebox{\textwidth}{!}{
\begin{tabular}{lll}
First DMA (Average \#Clearings) & Second DMA (Average \#Clearings) & Correlation \\ \hline
    West Palm Beach-Ft. Pierce, FL (3.203) & Miami - Ft. Lauderdale, FL (5.763) & -0.678 \\
    San Francisco-Oakland-San Jose, CA (7.877) & Sacramento-Stockton-Modesto, CA (2.500) & -0.539 \\
    Milwaukee, WI (5.230) & Chicago, IL (8.100) & -0.537 \\
    Washington, DC (4.690) & Baltimore, MD (3.693) & -0.482 \\
    Tampa-St Petersburg-Sarasota, FL (4.800) & Orlando-Daytona Beach-Melbourne, FL (6.197) & -0.440 \\
    Greenville-Spartanburg, SC-Asheville, NC (2.603) & Charlotte, NC (4.613) & -0.298 \\
    Dayton, OH (4.197) & Columbus, OH (3.380) & -0.258 \\
    Greensboro-High Point-Winston Salem, NC (4.833) & Charlotte, NC (4.613) & -0.254 \\
    Pittsburgh, PA (2.840) & Cleveland-Akron, OH (3.603) & -0.248 \\
    Philadelphia, PA (8.310) & New York, NY (10.177) & -0.245 \\
    Raleigh-Durham, NC (5.440) & Birmingham, AL (2.277) & -0.239 \\
    Dayton, OH (4.197) & Cincinnati, OH (2.643) & -0.222 \\
    Hartford-New Haven, CT (2.087) & Boston, MA (7.150) & -0.198 \\
\end{tabular}}
\caption{\it Top negatively correlated DMA pairs in terms of number of clearings in sampled solutions.
	Results restricted to DMAs with an average of at least 2 clearings, and pairs where the
	correlation was -1.0 or smaller with significance $P \leq 0.01$.}
\label{tab:corrpos}
\end{table}

It is worth noting that there are also weaker but still significant 
{\em positively\/} correlated DMA pairs, which are more difficult to understand
intuitively but still potentially meaningful daisy-chain effects.
For instance, the most strongly positively correlated pair is
Miami - Ft. Lauderdale, FL and
Charlotte, NC whose sample correlation is
0.191 and statistically significant.
While this magnitude is considerably smaller than those for the top negatively
correlated pairs, such positive correlations may
capture longer-distance or ``second order'' daisy-chain effects. 
One example would be subtle ``alternation'' effects --- 
for instance, if DMA $A$ is to the north of DMA $B$,
which in turn lies north of DMA $C$, then $A$ and $B$, and $B$ and $C$, will be negatively correlated; but
$A$ and $C$ may be positively correlated, because when there is more clearing in $A$,
less is required in $B$, but that necessitates clearing more in $C$. 

\subsection{Costs of Domain Constraints and Clearing Target at the DMA Level}

We can also perform DMA-level analysis with domain constraints and higher clearing targets. To clear 84 MHz with border 
constraints, one must now clear about 275 broadcasters nationwide. Thus, more broadcasters must now 
be cleared in each DMA. For instance, New York City now requires about 12 to be cleared on average, 
compared to 10 without domain constraints. The rankings also change in expected ways; namely, border cities such as Buffalo 
and San Diego now rank much higher. Both of these cities required less than 1 clearing on 
average without border constraints, but require more than 4 with the constraints. The two maps
in Figure~\ref{fig:map} illustrate the increase and redistribution of clearings required by adding domain constraints. 

As we increase the clearing target, the number of broadcasters that must be cleared on a per-DMA basis grows modestly. 
In general, across a wide range of clearing targets, there is an approximate ``5 in 15'' phenomenon ---
that is, it is necessary to clear more than about 5 broadcasters only in roughly the top 15 markets. 
To clear 96 MHz instead of 84 MHz, a full additional broadcaster needs to be cleared in just the top markets. 
In markets ranking below 10, the difference drops to about half a broadcaster.
The approximate cost, regardless of clearing target, of incorporating the border constraints is about 2 broadcasters per DMA,
again limited only to the largest DMAs. In lower ranked markets, the border constraints have little effect.

\begin{figure}[]
\centering
\subfigure[Sorted differences between amount of clearing required per DMA for 96 MHz vs. 84 MHz clearing target.]{\includegraphics[width=0.75\textwidth]{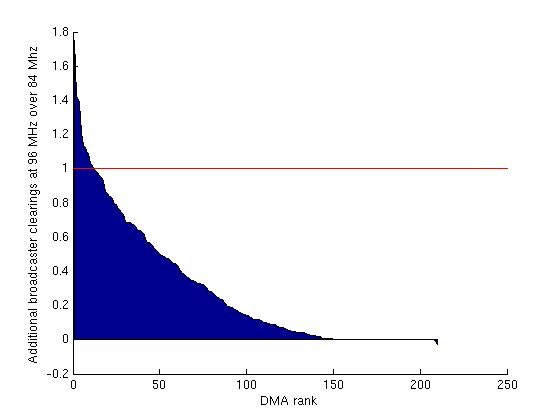}} \quad
\subfigure[Sorted differences between amount of clearing required per DMA with border constraints and without, at 84 MHz
clearing target.]{\includegraphics[width=0.75\textwidth]{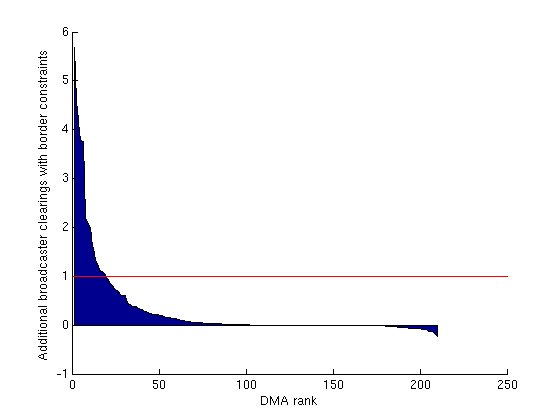}}
\caption{\it Effect of a higher clearing target or border constraints on the number of broadcasters clearings required per DMA. Both plots rank DMAs according to 
	the quantity being plotted. The horizontal red lines mark 1 additional broadcaster clearing.}
\label{fig:dmarank}
\end{figure}

Figure \ref{fig:dmarank} highlights how rapidly the costs of either a greater clearing target or 
border constraints diminishes with DMA rank. 
Figure \ref{fig:dmarank}(a)
shows the additional
number of broadcaster clearings required per DMA for a 96 MHz clearing target compared to an 84 MHz target.
Figure \ref{fig:dmarank}(b)
shows the differences in the amount of clearing required at 84 MHz between border 
constraints and no border constraints. In both cases, the fraction of DMAs where there is more than a 
1-surrender difference is very small. Thus, it seems that clearing 96 MHz is not qualitatively more difficult than 84 MHz, 
and that even the currently conservative border constraints significantly impact only a small number of markets. Note that in both plots,
there are a small number of DMAs with negative values, meaning that the average number of clearings in those DMAs actually
became {\em smaller\/} under the more challenging condition (higher clearer target or adding border constraints). In most
cases, these are DMAs that require very little clearing to begin with, and thus this reduction is likely simply due
to sampling error. 

\subsection{Minimum Number of Participating Markets}

We also studied the question of the minimum  number of markets whose participation in the auction is required to achieve a designated clearing target --- in other words,
what is the minimum number of DMAs in which {\em any\/} non-zero  amount of clearing is needed?
Using our sampled solutions, we can compute the average number of DMAs that require any 
amount of clearing, and also the minimum number 
of such DMAs observed in the sample. For 84 MHz, without domain constraints, the 
average was 73.9 DMAs, with a standard deviation of 3.2, and the minimum observed was 63. 
However, since this is just the observation in 300 sampled solutions, it leaves open the 
possibility that the true minimum might be less than 63. To determine this, we 
encode Boolean formulae that specify a maximum number $x$ of DMAs in which 
any broadcasters are cleared, and reduce $x$ until infeasibility occurs. The 
results demonstrate that there exists a solution in 
which only 43 DMAs require any clearing, and it is impossible to reduce the number further. 
Table \ref{tab:dmaclear} gives results of this form for various clearing targets, all ignoring 
domain constraints. Each number increases by about 10 broadcasters when these constraints are taken into account.

\begin{table}
\centering
\resizebox{\textwidth}{!}{
\begin{tabular}{llll}
Clearing Target & Average DMAs with Clearing & Standard Deviation & Minimum Possible \\ \hline
60 & 42.5900 &  2.9264 &  24 \\
66 & 46.1500 &  2.5775 &  27 \\
72 & 55.0200 &  2.8673 &  30 \\
78 & 60.4900 &  3.2440 &  38 \\
84 & 73.9800 &  3.2818 &  43 \\
90 & 80.8800 &  3.5310 &  58 \\
96 & 90.5800 &  3.2423 &  66 \\
102 & 102.3300 & 2.9782 & 68\\
108 & 109.0000 & 2.9162 & 77 \\
114 & 115.1800 & 3.3975 & 96 \\
120 & 120.4565 & 2.8915 & 106
\end{tabular}}
\caption{\it For each clearing target, and ignoring domain constraints, the average number of DMAs in which some clearing occurs, the standard deviation of this value, and the absolute minimum number of DMAs in which some clearing is needed to achieve feasibility. }
\label{tab:dmaclear}
\end{table}

\subsection{Clearing Buffers and Solution Diversity}

We now elaborate on our choice in the analyses above to allow the nationwide number of clearings to be slightly larger
than the absolute minimum necessary.
More precisely, in Table~\ref{tab:dma} and similar experiments, we enforced that the total number of broadcaster 
clearings be below the feasibility minimum plus a small 10-broadcaster ``buffer''. Thus,  
while the actual feasibility minimum for 84 MHz without domain constraints is 196 clearings, we allowed up to 
206 broadcasters to be cleared. This has the effect of enlarging the space of feasible solutions, and therefore 
generates a more varied distribution over still near-optimal solutions. The effect of the size of this 
buffer on the size of the solution space may be of independent interest, 
and  is shown in Table \ref{tab:buffers}. As the size of the 
buffer increases, the number of unique solutions found in our sampled set of 300 solutions increases, 
suggesting a more varied and less constrained solution set. Indeed, if we let $P$ be the distribution over
solutions generated by PicoSAT, we can perform a ``missing mass'' estimate of the total weight
under $P$ of solutions that do not appear in our sample; we use the standard estimate that divides
the number of solutions that only appear 
once in our sample by the total number of draws. \footnote{Gale, William A. ``Good-Turing Smoothing Without Tears''. Journal of Quantitative Linguistics 2: 217-237, 1995.}
As the table shows,
these estimates illustrate that sacrificing a small amount of optimality yields a much greater number of solutions,
since a buffer of only 10 clearings already increases the missing mass from 39\% to 71\%.
\footnote{Note that we expect the missing mass to be spread fairly evenly over a large number of unseen solutions,
as opposed to concentrated on small number, since in the latter case we would have seen these high-probability
solutions in the sample itself.}

Just because the number of solutions increases, however, does not mean that the ``diversity'' of solutions increases as well. 
One measure of diversity is the 
difference between the set of broadcasters cleared in different solutions. 
In fact, by this measure, regardless of buffer size, the diversity is relatively constant. To quantify this, for a pair of solutions, let $x$ denote the number of broadcasters in the two assignments for which the clearing status differs (i.e., the broadcaster was cleared in one solution but not the other). Let $y$ denote the maximum possible value of $x$, which is the sum of the number of broadcasters cleared in the two assignments. 
Then define the ``distance'' between these two assignments to be $x/y$. 
\footnote{
More formally, if $S_1$ is the set of broadcasters cleared in one solution
and $S_2$ the set cleared in another, we define the distance between the two
solutions to be $|(S_1 \cup S_2) - (S_1 \cap S_2)|/|S_1 \cup S_2|$.} 
Empirically, on average, 
the distance or diversity is about 0.5. In other words, about 50\% of the possible broadcaster clearing differences that could occur actually do. This is quite high, and suggests that the solution spaces are indeed fairly diverse.

The above analysis does not indicate where this diversity occurs geographically. To measure this, we repeat the analysis at a DMA level, and therefore determine the average solution diversity for each DMA. Intuitively, a DMA with high diversity is one in which there is a lot of flexibility 
regarding which broadcasters can be cleared to reach the target. The DMAs with the highest diversity are shown in Table \ref{tab:diversity}, and form an interesting geographic mix. Some major DMAs such as Los Angeles and Philadelphia appear, as well as several smaller ones 
such as Denver, Syracuse, and Tulsa. 

\begin{table}
\centering
\begin{tabular}{l l l l}
Buffer & Unique Solns & Solns Appearing Once & Missing Mass Estimate \\ \hline
0 & 190 & 117 & 39.0 \% \\
5 & 241 & 195 & 65.0\% \\
10 & 255 & 214 & 71.3\% \\
20 & 261 & 226 & 75.0\%
\centering
\end{tabular}
\caption{\it Effect of broadcaster ``buffer'' on size of solution space at 84 MHz clearing target.}
\label{tab:buffers}
\end{table}

\begin{table}
\centering
\begin{tabular}{l l l}
Rank & DMA & Diversity \\ \hline
1 & Denver, CO & 0.8380 \\
2 & Ft. Smith-Fayetteville-Springdale-Rogers, AR & 0.8075 \\
3 & Dallas-Ft. Worth, TX & 0.7835 \\
4 & Pittsburgh, PA & 0.7525 \\
5 & Detroit, MI & 0.6793 \\
6 & San Antonio, TX & 0.6787 \\
7 & West Palm Beach-Ft. Pierce, FL & 0.6724 \\
8 & Syracuse, NY & 0.6658 \\
9 & Sacramento-Stockton-Modesto, CA & 0.6658 \\
10 & Los Angeles, CA & 0.6651 \\
11 & San Francisco-Oakland-San Jose, CA & 0.6625 \\
12 & Cleveland-Akron, OH & 0.6598 \\
13 & Philadelphia, PA & 0.6486 \\
14 & Providence, RI-New Bedford, MA & 0.6350 \\
15 & Miami - Ft. Lauderdale, FL & 0.6294 \\
16 & Hartford-New Haven, CT & 0.6289 \\
17 & Indianapolis, IN & 0.6233 \\
18 & Milwaukee, WI & 0.6184 \\
19 & Tulsa, OK & 0.6072 \\
20 & Atlanta, GA & 0.5918 \\
\centering
\end{tabular}
\caption{\it DMAs with the highest solution diversity.}
\label{tab:diversity}
\end{table}

\begin{figure}[]
\centering
\includegraphics[width=0.75\textwidth]{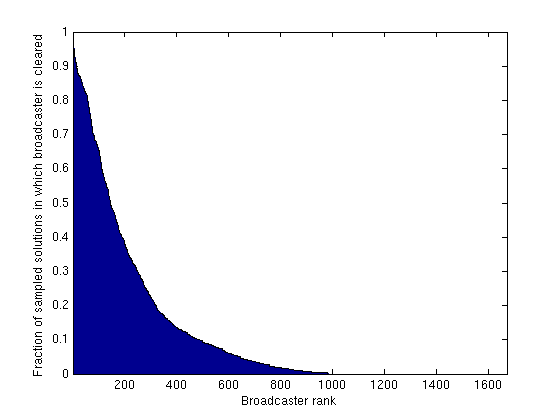}
\caption{\it Sorted fractions of 300 sampled feasible solutions in which each broadcaster is cleared at 84 MHz. 
}
\label{fig:bcdifficulty}
\end{figure}

\subsection{Constraints at the Broadcaster Level}

Finally, while so far we have reported on the absolute constraints on feasible solutions at the national and
per-DMA levels, each sampled solution of course commits to specific choices of which individual broadcasters
are cleared and repacked. While we shall refrain from providing detailed analyses and identities at the broadcaster level,
it is instructive to investigate the frequency with which each broadcaster is cleared in the effort to reach 
a nationwide target in our sample of 300 solutions. Figure~\ref{fig:bcdifficulty} shows the fraction of samples
in which each broadcaster is cleared, after sorting broadcasters by the same quantity.
There are a few broadcasters who are cleared almost always -- one is
cleared in 97\% of samples, and there are 3 others who are cleared in 96\%.
These are broadcasters that, at least with respect to the distribution of solutions generated by PicoSAT,
present a relatively challenging interference profile that leads to frequent clearing. 
However,
the dropoff is steep, 
with only 14 broadcasters who were cleared in more than 90\% of solutions. 

Note also that there are 1672 broadcasters
in total, and a very large fraction are not cleared in any solution. Thus it appears that the hard constraints
on regional clearing emerge only at the DMA level and above, and not at the resolution of individual broadcasters.

\section{Incorporating Broadcaster Participation Models}

All the analyses presented so far assume that every broadcaster participates in the auction and is willing to be cleared off the air. 
This assumption allows us to identify the absolute and purely structural constraints on the solution space, at the national, DMA, and broadcaster levels. 
However, this approach also ignores any assumptions, beliefs or information we have about the relative likelihood that various broadcasters will participate in the auction. 
To analyze what ``might'' occur, rather than what ``must'', we consider a different type of analysis that incorporates models of broadcaster participation. In doing so, we do not seek to make any judgment about who should participate. We simply seek to explore repacking feasibility given specific and varying assumption sets for the purpose of understanding likely outcomes.
Our models are expressed as joint probability distributions over 
broadcaster participation decisions. 
We can design this distribution to reflect various assumptions about the probabilities of, and correlations between, broadcaster participation decisions, and therefore
obtain results that more closely model reality under our beliefs. 
Rather than extracting information at the DMA level, this analysis focuses on groups of broadcasters sharing similar properties, such as network affiliate status.

\subsection{Models and Methodology}

We now outline the nature and purpose of each of the successive models
we examine. Each of these models specifies a 
parameterized class of joint probability distributions $P(\vec{x})$ over the
vector $\vec{x}$ of broadcaster participation decisions.
Because the technical specification of the more complex models is
somewhat mathematical, we defer these details to the Technical Appendix in Section~\ref{sec:part-models}.

\begin{itemize}
\item {\em Random Broadcasters Model.\/}
This is our simplest model, which has a single parameter $\alpha$ representing the
probability each broadcaster chooses {\em not\/} to participate in the auction (or exits without clearing), 
and is therefore
repacked; 
\footnote{Again, in keeping with our desire to avoid pricing assumptions or other mechanism-specific
details, we deliberately conflate non-participation with participating but exiting at some point,
and participation with eventual clearing at an unspecified price.}
thus with probability $1 - \alpha$, the broadcaster is cleared. Each broadcaster makes
an independent decision of this type, hence the name of the model. In this model there will be
no correlations between the participation decisions of different broadcasters. By choosing or varying the value of
$\alpha$, we can model more or less (probabilistic) participation in the overall population of broadcasters.
Obviously by making $\alpha$ larger,
we expect to make the probability of reaching any fixed clearing target smaller; the question is what the
form and rate of this relationship takes.
\item {\em Random Affiliates Model.\/}
The assumption that all broadcasters make their decisions independently is obviously a vast
oversimplification, so we also consider models allowing correlations between the decisions of
related subsets of broadcasters. 
The random affiliates model represents a scenario in which there are separate and independent 
events controlling the participation of each network affiliate group (ABC, NBC, FOX, CBS, PBS \footnote{We include PBS because these broadcasters are likely to have common influences affecting their participation decisions, not because we believe their motives are similar to those of the other four networks listed above.}), capturing the notion that
(for example) internal corporate decisions or influences at each affiliate may strongly determine the participation
decisions of its constituents.
In this model, if the
event for a particular affiliate group occurs, which happens with probability $\alpha$, then 
{\em all members\/} decline to participate in the auction. If the event does not occur, 
then all members in the affiliate group participate. All non-affiliates still make an independent decision, again with probability $\alpha$
of non-participation. Note that for any fixed value of $\alpha$, the
{\em marginal\/} (that is, isolated) probability of any particular broadcaster not participating will be
$\alpha$, exactly as in the random broadcasters model. However, now very strong correlations
are present in the joint distribution --- namely, knowing the participation status of any broadcaster
in an affiliate group immediately determines the same status for all others in the group. We recognize that this model still 
oversimplifies the complex decision-making broadcasters will undertake in determining whether to participate or not, but we believe this offers 
a refinement of the random broadcasters models that is useful in determining the effects of
affiliate correlations.
\item {\em Correlated Affiliates Model.\/}
This model captures an even richer scenario, in which  
there is not only correlation between the broadcasters within an affiliate, but also
across affiliates themselves --- thus rather than no correlations at all (random broadcasters model),
or only one ``layer'' of correlation (random affiliates model), now we have two layers of correlation ---
one between affiliates, and one between broadcasters within an affiliate. 
The top-level correlation captures the outcome of events that might strongly influence the participation decisions of all the network affiliates.
Again, the model is designed so that the marginal probability of non-participation is a fixed
value $\alpha$ across all broadcasters, and can be varied to examine different assumptions on
the background rate of participation nationwide. Non-affiliates still remain independent of all other broadcasters
with probability $\alpha$ of non-participation.
(See Section~\ref{sec:part-models} for technical specifications.)
This permits direct comparison to the models
above at a given marginal rate $\alpha$, allowing us to isolate the effects of the correlations alone.
\end{itemize}

Although in the descriptions above we refer to probabilistic events and correlations,
we can informally and alternately view the models
as capturing uncertainty (as opposed to randomness) and shared traits across groups of broadcasters. 
For instance, in the random broadcasters model, we can view $\alpha$ as representing an assumption about
the fraction of the $n$ broadcasters that will not participate, and the randomness as representing agnosticism
regarding exactly which subset of size roughly $\alpha n$ will not participate (all equally likely).
Similarly, in the random affiliates model the shared random event can be viewed as a crude proxy for the
fact that affiliates of the same network share properties or traits that make them more likely to behave
in unison.

Given any of the models above (or any other joint probability distribution $P(\vec{x})$), and fixing a clearing target, we can 
calculate the probability that a feasible solution can be reached. To do so, we run an experiment 
consisting of multiple trials, each of which represents a potential outcome of the auction. In each 
trial, we draw from the distribution $P(\vec{x})$ to make a random decision about which 
broadcasters will choose to participate. If a broadcaster participates in the auction, then it can be 
cleared. If a broadcaster does not participate, then it must be repacked. We then use 
PicoSAT to determine whether it is possible to repack the set of non-participating broadcasters on the available 
channels, subject to the constraints. By repeating enough such trials, we can approximate the probability of success at different
clearing targets. The 
formal description of the algorithm is given in the Technical Appendix in Section \ref{sec:pico-algo}.

Before presenting the findings, we wish to emphasize that there is no special status to any of the
models examined --- they were chosen for their balance between capturing potential real-world
assumptions about participation rates and broadcaster decision correlations, and mathematical simplicity
(e.g. having few free parameters). Obviously all manner of hybrids of these models could be considered,
as well as entirely different models accounting for potentially relevant properties of broadcasters such
as geographic location, population density, industry events and views, and so on. Our goal here is not
to present definitive findings, but to illustrate the kinds of outcomes different models lead to.

\subsection{Results: Clearing Probabilities for the Participation Models}

\begin{figure}
\centering
\subfigure[Random broadcasters model: each broadcaster has an $\alpha$ probability of non-participation.]{\includegraphics[width=0.45\textwidth]{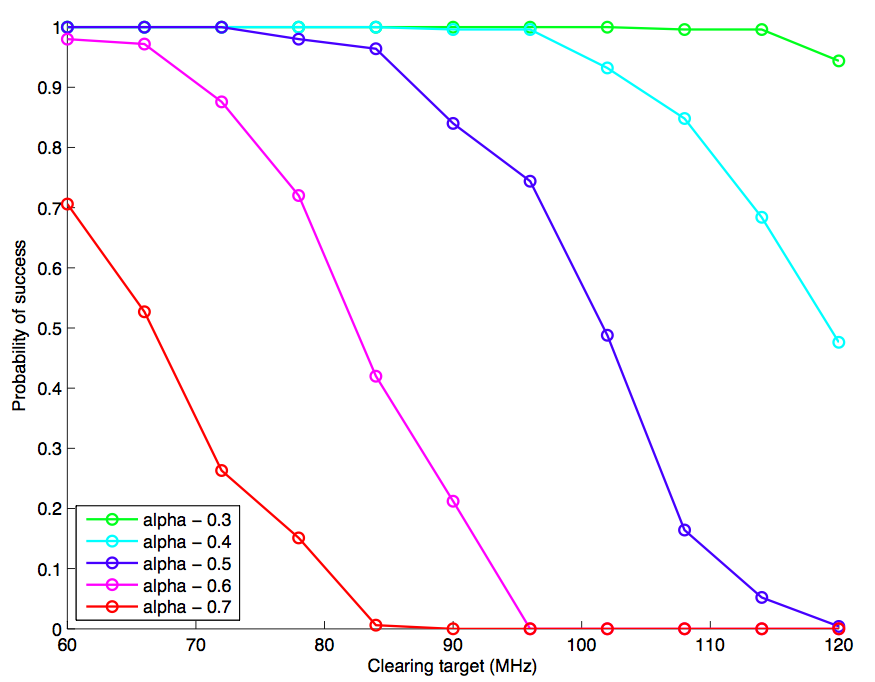}} \quad
\subfigure[Random affiliates model: each affiliate group has $\alpha$ probability of complete non-participation and $1-\alpha$ probability of complete participation.]{\includegraphics[width=0.45\textwidth]{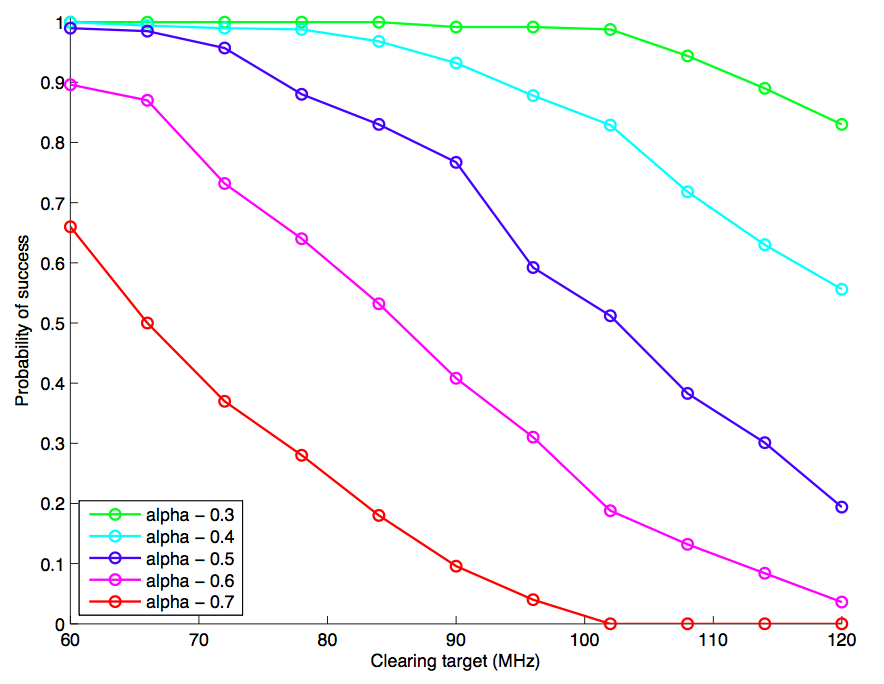}} \\
\subfigure[Correlated affiliates model: similar to random affiliates model, but participation 
of each group is correlated.]{\includegraphics[width=0.45\textwidth]{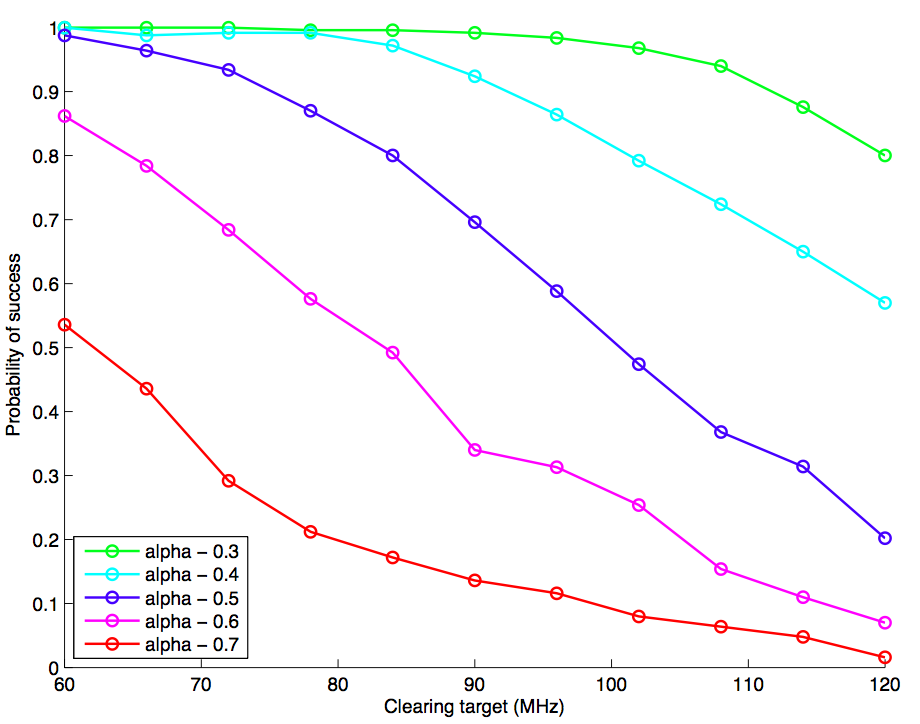}} \quad
\subfigure[All three models, with alpha fixed at 0.6.]{\includegraphics[width=0.45\textwidth]{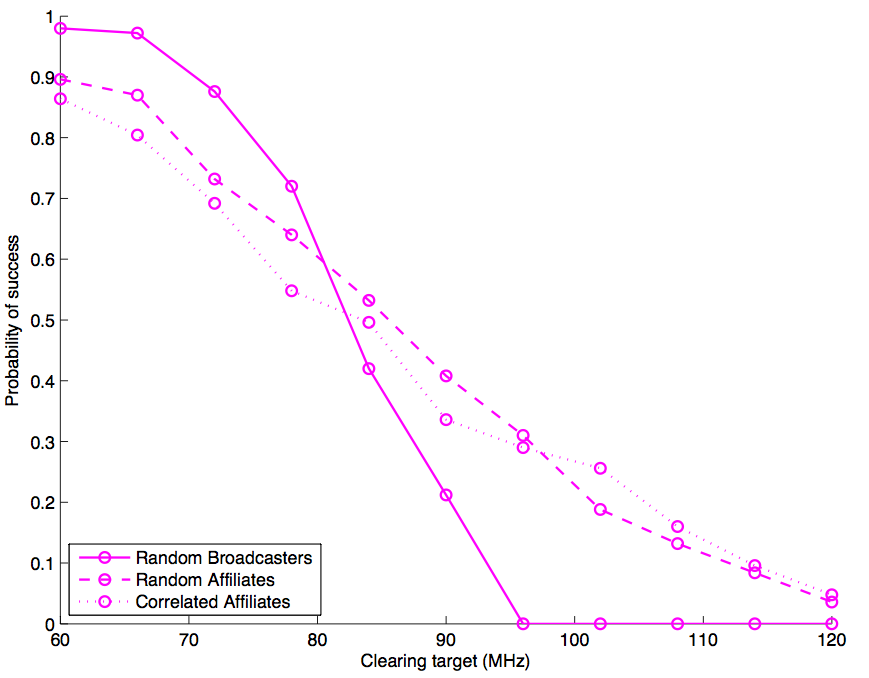}}
\caption{\it Success probabilities for the first three participation models, as a function of clearing target, and ignoring domain constraints. 
Each model represents successively greater correlation between broadcaster participation decisions. As correlation increases, lower clearing targets are harder to reach, and higher targets are easier to reach. This trend is evident in subfigure (d), which plots all three models for alpha = 0.6. The random broadcasters model has the highest success probability at 60 MHz and the lowest at 120 MHz, whereas the correlated affiliates model has the lowest success probability at 60 MHz and the highest at 120 MHz.}
\label{fig:success}
\end{figure}

In Figure \ref{fig:success}, we plot the probability of success for the first three participation 
models (all without border constraints) as a function of clearing target. One 
immediately evident trend is that increasing the correlation between broadcaster participation probabilities makes 
it more difficult to reach lower clearing targets, but easier to reach higher clearing targets.
In general,
the curves for the correlated models start lower but decay more gradually than for random broadcasters.
Thus, for any fixed value $\alpha$ for the marginal probability of non-participation or repacking across the three models,
the random affiliates and 
correlated affiliates models are more pessimistic (less likely to clear a given target)
than the baseline random broadcasters model in the 60 - 78 MHz range;
whereas at the higher targets in the
78 - 120 MHz range, and especially for larger $\alpha$,
the correlated models are more optimistic.

We believe these asymmetric effects of correlation are largely due to the nature of the variance in the differing models.
In the random broadcasters model, since each broadcaster decision is an independent coin flip, the fraction of broadcasters
nationwide and even in broad geographic regions (such as the highly congested Northeast Corridor) will be sharply
concentrated around its expectation $\alpha$ --- there is relatively little chance of either ``getting lucky" and clearing
unexpectedly large numbers of broadcasters, or ``getting unlucky'' and clearing 
unexpectedly small numbers of broadcasters. Relative to the correlated models, the random broadcasters
curves exhibit threshold behavior, in which
clearing probabilities remain high for a time before dropping sharply.
However, clearing the higher targets requires getting lucky, and the correlated models permit this
when the correlating events or hidden variables assume values that cause much larger than average participation
from affiliate broadcasters. The flip side of this is that much smaller than average participation is also possible.
In statistical terms, even though the three models have the same average participation rate $\alpha$, the correlated
models have much higher variance --- achieving an average rate of $\alpha$ by mixing cases of much higher and lower participation. 
One can even see this effect between the random affiliates and correlated affiliates models at $\alpha = 0.7$
(red curves in Figure~\ref{fig:success}(b) and (c)) --- the former is more likely to clear low targets, the latter more likely to clear
high targets. Again, this is due to the correlated affiliates model having an additional layer of correlation, thus
increasing variance.

\begin{figure}
\centering
\subfigure{\includegraphics[width=0.5\textwidth]{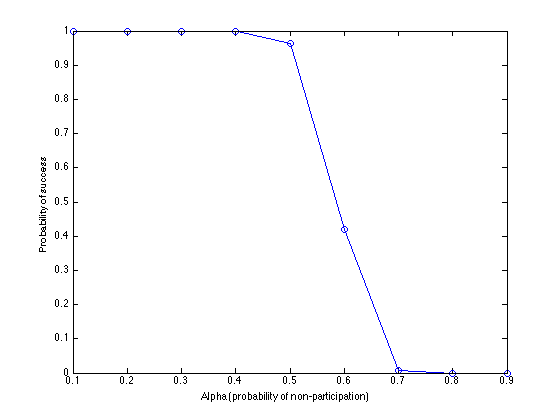}}
\caption{\it Success probability for a clearing target of 84 MHz in the random broadcasters model, as a function of $\alpha$. The curve exhibits a steep 
threshold phenomenon, with the tipping point at about $\alpha = 0.6$.}
\label{fig:threshold}
\end{figure}

The sharp concentration or low variance of the random broadcasters model is highlighted by
Figure \ref{fig:threshold}, which plots the probability of success for this 
model as a function of $\alpha$ at the fixed target of 84 MHz. A 
strong threshold phenomenon can be observed, with an inflection point at around $\alpha = 0.6$. At 
lower values of $\alpha$, success is almost guaranteed, whereas at higher values, failure is nearly certain. 

\begin{figure}[]
\centering
\subfigure[Without domain constraints]{\includegraphics[width=0.45\textwidth]{indep}}\qquad
\subfigure[With domain constraints]{\includegraphics[width=0.45\textwidth]{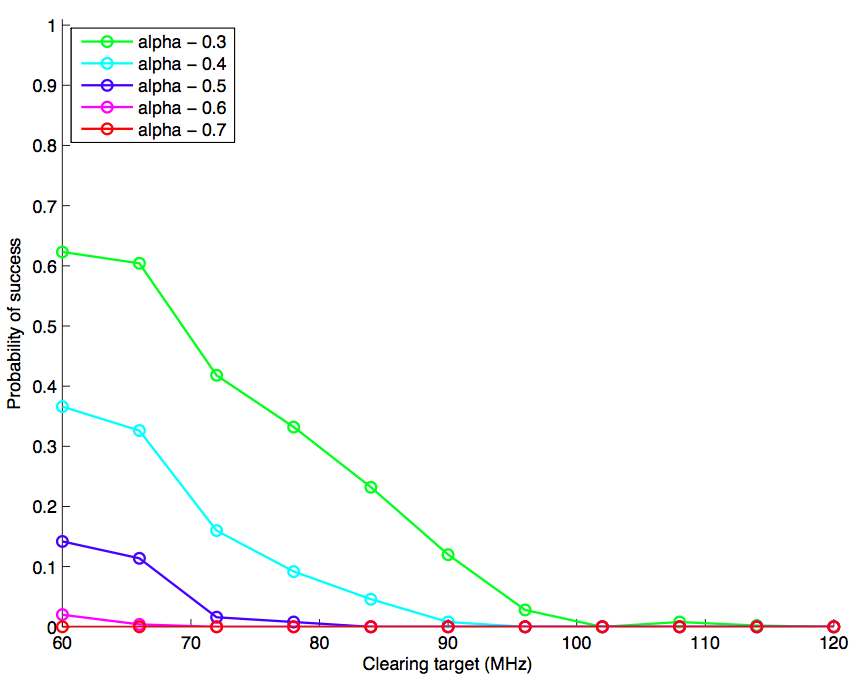}}\qquad
\caption{\it Success probabilities for the random broadcasters model, with and without domain constraints.}
\label{fig:indep-domain}
\end{figure}

Note that all results discussed so far ignore domain constraints. 
Figure \ref{fig:indep-domain} provides a 
comparison of the random broadcasters model success probabilities, both with and without domain constraints. A similar effect can be observed in all participation models. 
As can be seen, the probability of success is severely 
reduced when domain constraints are taken into account. 
This reduction is largely caused by the existence of broadcasters in the border regions 
that can only be repacked onto a few channels, if any, due to the current treaties 
with Canada and Mexico. Thus, the non-participation of these broadcasters severely restricts the 
set of feasible solutions. 

Note that the ``might'' experiments discussed in this section present a rather more pessimistic 
view of the domain constraints than the ``must'' experiments of the previous section. Recall 
that in the ``must'' experiments, we assume that all broadcasters are willing to participate in 
the auction.
As a result, we can 
assume the broadcasters in particularly tricky border regions will be 
cleared off the air. However, in the ``might'' experiments, we have no control 
over which broadcasters participate. The decisions are determined for us by a 
random draw from a probability distribution. Thus, it is quite likely that some of the 
border broadcasters will end up choosing not to participate, thereby making the problem infeasible.

\subsection{Causes of Infeasibility}
\label{sec:blockingset}

Recall that a repacking problem is feasible if, given a clearing target $M$ and a set $S$ non-participating broadcasters, all broadcasters in 
$S$ can be repacked onto the $c = 38 - \lfloor M/6 \rfloor$ remaining channels. Consider a set of $c+1$ broadcasters with the property that each pair $i,j$ in the set cannot be repacked onto the same channel. Such a set corresponds to a \emph{clique} in the co-channel interference graph. 
If all broadcasters in this set choose not to participate, then there is simply no way to repack them all onto $c$ channels. Thus, given a clearing target $M$ (or equivalently, $c$ available channels) and a set $S$ of non-participating broadcasters, if the set $S$ contains any co-channel clique of size $c+1$, we immediately know the problem is infeasible. We call such a clique a \emph{blocking clique}.

Note that a problem can be infeasible for other reasons as well. In particular, the cause might involve the interplay of both co-channel and adjacent-channel interference constraints. 
However, an interesting empirical observation is that, at least when domain constraints are ignored, the most common cause of infeasibility is a blocking clique.
For instance, for the random broadcasters, random affiliates, and correlated affiliates participation models discussed above,
the percentage of infeasible cases that are due to the existence of a blocking clique ranges
from about 75\% to 85\% at a 60 MHz clearing target, and increases to
almost 100\% for all three models at clearing targets above 90 MHz.

This observation has algorithmic and other benefits in determining feasibility.
The fact that infeasibility can usually be traced to a blocking clique allows us to bypass the use of a SAT 
or ILP solver entirely, by simply first checking whether the set of non-participating broadcasters contains a blocking clique.
We can compute a set of cliques ahead of time, so this check can be done very rapidly. A single trial with PicoSAT takes on average 30-40 seconds, and can go on indefinitely if a timeout is not used. A single trial of the alternate methodology is almost instantaneous. The algorithm is described formally in the Technical Appendix in Section \ref{sec:clique-algo}.

Note that there might be exponentially many cliques in a graph, so finding all of them is computationally intractable.
Thus, we can instead use a greedy method to enumerate as many as possible. As evidence that this heuristic has near-optimal 
performance, we note that the percentages of infeasibilities attributed to cliques given above were indeed found via the greedy method. 


\subsection{Degree of Infeasibility}

Thus far we have been treating feasibility as binary, i.e. either our clearing target can be reached nationwide, or not. 
However, there are many reasons why failure can occur, and some are more severe than others. For instance, suppose we have a 
clearing target of 84 MHz, which leaves 24 channels open. If the set of non-participating broadcasters contains exactly one blocking clique of size 25, then we have infeasibility. Yet this infeasibility is quite weak, in the sense that if we 
are content to clear 78 MHz in the region containing the blocking clique, we can reach our clearing target of 84 MHz in the rest of the country. 
On the other hand, if the set of non-participating broadcasters contains multiple blocking cliques across the country, then this infeasibility is more problematic.

One benefit of using the clique-based methodology discussed in the previous section is that we always know the cause of failure. Thus, given a clearing 
target and participation model, we can measure how much of the country will typically be affected by the event of infeasibility. 
Let $z$ indicate the average number of broadcasters that appear in any blocking clique when infeasibility occurs. For instance, if infeasibility 
occurs due to the existence of a single blocking clique, then $z$ equals the size of that blocking clique. If infeasibility occurs due to 
the existence of multiple blocking clique, then $z$ equals the size of the union of these blocking cliques. Intuitively, the larger 
the value of $z$, the greater the fraction of the country affected, and therefore the ``worse'' the infeasibility. 

\begin{figure}[]
\centering
\includegraphics[width=0.65\textwidth]{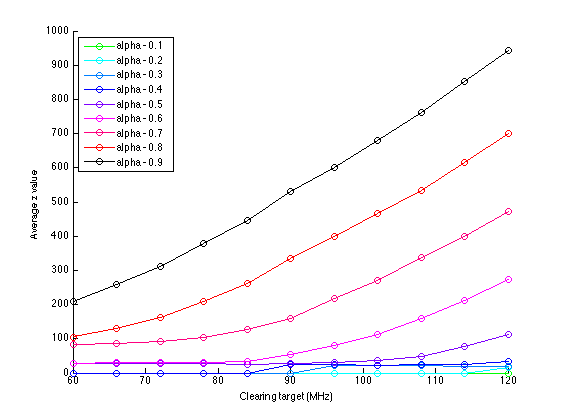}
\caption{\it Average $z$ values for various clearing targets in the random broadcasters model. The average $z$ value increases as a function of both clearing target and $\alpha$ (i.e. as the repacking problem becomes more constrained).}
\label{fig:zvals}
\end{figure}

In Figure \ref{fig:zvals}, we plot the average $z$ value for various clearing targets in the random broadcasters model. Note that we are conditioning on the event of infeasibility; i.e. the plots only illustrate trials in which failure occurred. In each plot, the average $z$ value increases as a function of both clearing target and $\alpha$. A value in the 0 - 100 range suggests that the infeasibility is quite local, and only
affects a small region of the country. A value above 300 suggests that many different areas of the country are affected, though not all are significantly impaired. Finally, a value above 500 indicates that the failure is both widespread and severe.
The analogous figure looks numerically quite similar for the random affiliates and correlated affiliates models. 

\begin{figure}[]
\centering
\subfigure[$\alpha = 0.6, z=52$]{\includegraphics[width=0.45\textwidth]{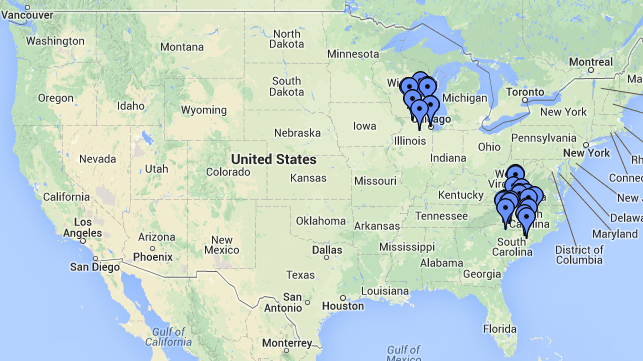}}\qquad
\subfigure[$\alpha = 0.7, z=106$]{\includegraphics[width=0.45\textwidth]{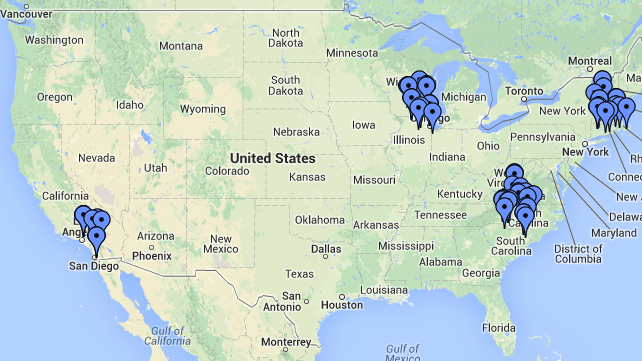}}\qquad
\subfigure[$\alpha = 0.8, z=222$]{\includegraphics[width=0.45\textwidth]{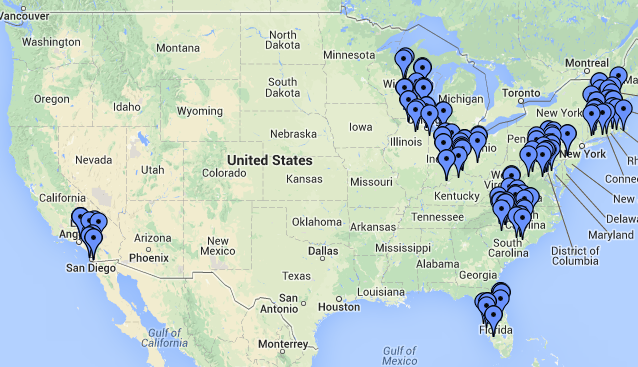}}\qquad
\subfigure[$\alpha = 0.9, z=450$]{\includegraphics[width=0.45\textwidth]{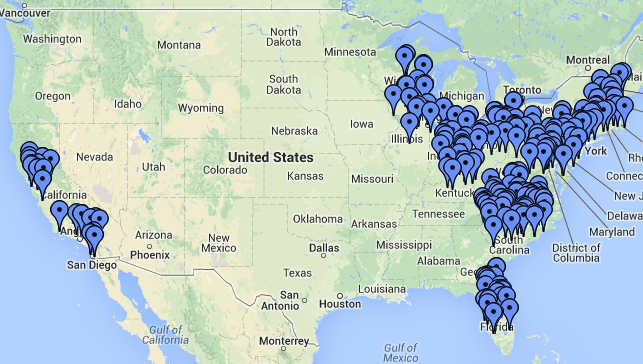}}\qquad

\caption{\it Illustration of different $z$ values for a single trial run of the random broadcasters model for a clearing target of 84 MHz. As $\alpha$ increases, $z$ does as well, and infeasibility affects larger portions of the country.}

\label{fig:bsets}
\end{figure}

Figure \ref{fig:bsets} 
provides a geographic illustration of the degree of infeasibility $z$ at four increasingly large values
of $\alpha$, and was generated by running the random broadcasters model at an 84 MHz clearing target with
``shared randomness'', meaning that any non-participation choices made at one value of $\alpha$ are carried over to the next
higher value of $\alpha$. 
For this particular run, at $\alpha = 0.6$ we have $z = 52$, and infeasibility is due to two isolated blocking cliques only. 
As we increase $\alpha$ to $0.9$, the value of $z$ climbs to $450$, and the infeasibility 
extends to blocking cliques all over the country, including California and Florida. 
The basic point is that even conditioning on infeasibility, there is a difference between small and large non-participation rates,
in the sense that the {\em extent\/} of infeasibility is quantifiably worse in the latter than the former.
We note that the particular blocking cliques shown are of course specific to this particular random trial, and different
runs will lead to different specific problem areas.

\section{Conclusions and Future Work}

Our analyses have illustrated various properties of the solution space of the repacking problem. In particular, we have determined what ``must'' happen for the problem to be feasible, and what ``might'' happen under different probabilistic participation models. In the former category, our results highlight the effects of clearing target and domain constraints on the type, size, and diversity of solutions, at both a national and DMA level. In the latter category, we examined the effects of various assumptions about broadcaster participation, namely based on network affiliate status. Our hope is that the presentation of these results 
provides a quantitative framework in which to discuss and analyze the repacking problem. 
Moving forward, as more information about and new data for the auction comes to light, 
we plan to extend and refine our analyses and participation models.

\section{Technical Appendix}

\subsection{CNF Encoding}
\label{sec:cnf}

Each satisfiability instance takes the following parameters as input:

\begin{itemize}
\item (required) MHz clearing target $M$, or equivalently a channel clearing target $\lfloor M/6 \rfloor$, which is converted to a number $c = m - \lfloor M/6 \rfloor$ of channels available to repack on
\item (optional) Maximum number $b$ of broadcasters to clear nationwide
\item (optional) Maximum number $b'$ of broadcasters to clear from a specified DMA $z$
\item (optional) Maximum number $d$ of DMAs in which clearing is allowed to occur
\item (optional) Set $R$ of broadcasters that must be repacked
\end{itemize}

We express the interference and domain constraints in the form of binary matrices. We have a co-channel interference matrix $Co$ of size $n \times n$, where $Co(i,k) = 1$ encodes the constraint $A(i) \neq A(k)$. We have an adjacent-up-channel interference matrix $AdjUp$ of size $n \times n$, where $AdjUp(i,k) = 1$ encodes the constraint $A(i) \neq A(k) + 1$, and we have an adjacent-down-channel interference matrix $AdjDown$ of size $n \times n$, where $AdjDown(i,k) = 1$ encodes the constraint $A(i) \neq A(k) - 1$. Finally, we have a domain matrix $Domain$ of size $n \times m$, where $Domain(i, j) = 1$ encodes the constraint $A(i) \neq j$.

To solve the satisfiability problem, we encode the problem as a Boolean formula in conjunctive normal form (CNF). Let $B = \{1, \ldots, n\}$, $C = \{0, \ldots, c\}$. Let $x_{ij}$ for $i \in B, j \in C$ be 1 iff broadcaster $i$ is assigned to channel $j$. Let $x_{i0}$ for $i \in B$ be 1 iff broadcaster $i$ is not assigned to any channel. Let $D_j \subseteq B$ for $j \subseteq \{1, \ldots, 210\}$ denote the set of broadcasters belonging to DMA $j$. Let $y_j$ for $j \in \{1, \ldots, 210\}$ be 1 iff no broadcasters are cleared from DMA $j$. The boolean formula is as follows:
\begin{align}
& \bigvee_{j \in C} x_{ij} & \forall i \in B - R \\
& \bigvee_{j \in C-[0]} x_{ij} & \forall i \in R \\
& \bigwedge_{j, k \in C} \neg x_{ij} \vee \neg x_{ik} & \forall i \in B \\
& \neg x_{ij} \vee \neg x_{kj} & \forall i,k \in B \text{ s.t. } Co(i,k) = 1, \forall j \in C \\
& \neg x_{ij} \vee \neg x_{k(j+1)} & \forall i,k \in B \text{ s.t. } AdjUp(i,k) = 1, \forall j \in C \\
& \neg x_{ij} \vee \neg x_{k(j-1)} & \forall i,k \in B \text{ s.t. } AdjDown(i,k) = 1, \forall j \in C \\
& \neg x_{ij} & \forall i \in B, \forall j \in C \text{ s.t. } Domain(i,j) = 1 \\
& \texttt{at\_most\_true}(\{x_{i0} \mid i \in B\}, b) \\
& \texttt{at\_most\_true}(\{x_{i0} \mid i \in D_z\}, b') \\
& \neg x_{i0} \vee y_j & \forall i \in B, \text{ where } i \in D_j \\
& \bigvee_{i \in D_j} x_{i0} \vee \neg y_j & \forall j \in \{1, \ldots, 210\} \\
& \texttt{at\_most\_true}(\{y_{j} \mid j \in \{1, \ldots, 210\}\}, d) 
\end{align}

Lines 1, 2, and 3 ensure that each broadcaster is assigned to exactly 0 or 1 channels, unless the broadcaster is a member of $R$, in which case it must be assigned to exactly 1 channel. Lines 4, 5, and 6 encode the co- and adj- channel interference constraints, and line 7 encodes the domain constraints. The function \texttt{at\_most\_true}$(S, k)$ takes as input a set of variables $S$ and a number $k$ and returns a CNF encoding of the constraint that at most $k$ variables in $S$ can be true. The encoding we use is from the paper ``Towards an optimal CNF encoding of Boolean cardinality constraints'' by Carsten Sinz. \footnote{Sinz, Carsten. ``Towards an optimal CNF encoding of Boolean cardinality constraints.'' In Proc. of the 11th Intl. Conf. on Principles and Practice of Constraint Programming, pages 827-831. 2005.} Thus, line 8 encodes the constraint that at most $b$ broadcasters are cleared nationwide. Line 9 encodes the constraint that at most $b'$ broadcasters are cleared from DMA $j$. Lines 10, 11, and 12 encode the constraint that clearing occurs in at most $d$ DMAs.

\subsection{PicoSAT Algorithm}
\label{sec:pico-algo}

Algorithm \ref{alg:picosat} provides a formal description of our approach. The algorithm takes as input a joint distribution $P(\vec{x})$ and a clearing target $M$. Optional additional inputs are a maximum number of broadcasters to clear nationwide, a maximum number of broadcasters to clear from a particular DMA, and a maximum number of DMAs in which clearing is allowed to occur. The output is the probability $p$ that a feasible solution can be reached.

\begin{algorithm}
\caption{\it PicoSAT Algorithm}
\label{alg:picosat}
\begin{algorithmic}
\State $n = 1672$ \Comment{Number of broadcasters}
\State $m = 38$ \Comment{Number of channels}
\State $c = m - \lfloor M/6 \rfloor$ \Comment{Number of available channels}
\State $\text{infeasibilities} = 0$
\For {$i=1$ to $T$}
\State $x \leftarrow P(\vec{x})$ \Comment{Sample $x$ from $P(\vec{x})$}
\State $A \leftarrow \text{PicoSAT}(n, c, x)$ \Comment{Run PicoSAT}
\If {$\neg A$} \Comment{Infeasibility occurred}
\State $\text{infeasibilities} \gets \text{infeasibilities} + 1$
\EndIf
\EndFor
\State $p = 1 - \text{infeasibilities}/T$ \Comment{$p$ is probability of success}
\end{algorithmic}
\end{algorithm}

\subsection{Detailed Specifications of Broadcaster Participation Models}
\label{sec:part-models}

We now provide a more formal description of the broadcaster participation models. Recall that each model specifies a class of probability distributions over the vector of broadcaster participation decisions. The models introduce varying degrees of correlation between these decisions. The random broadcasters model has no such correlations, and is a simple product distribution in which each broadcaster has probability $\alpha$ of choosing not to participate in the auction. In the random affiliates and correlated affiliates models, we model correlations by introducing {\em hidden variables\/}
that do not represent broadcaster decisions, but instead might represent higher-level events
whose outcome could influence those decisions. 

In the random affiliates model, we have one hidden variable each for the networks ABC, CBS, FOX, NBC, and PBS. Each of these variables takes the value of 1 with probability $\alpha$. If a network variable has value 1, then all member broadcasters also have value 1, which indicates non-participation. If a network variable has value 0, then all member broadcasters also have value 0.

In the correlated affiliates model, we again have one hidden variable for each of the networks, and the effect of this variable is the same; i.e. its value propagates down to all members. We also have an additional top-level variable that influences all the second-level affiliate variables.
If the top-level variable takes a value of 1, which happens with $0.9$ probability, 
then each affiliate variable has $\frac{\alpha}{0.9}$ probability of taking value 1,
and $1-\frac{\alpha}{0.9}$ probability of taking value 0. If the top-level variable takes a value of 0, then 
each affiliate group has probability 1 of taking value 0. In the previous model, the probability that each non-affiliate
group chose not to participate was fixed; here, the top-level variable acts as a ``switch'' that either raises
or lowers this probability for each group simultaneously. Note that the probability of $0.9$ 
is chosen for mathematical reasons, and does not have a strong effect on the results observed in this model. 
All non-affiliates again make an independent decision.

\begin{figure} 
\begin{minipage}[]{0.45\linewidth}
\centering
\includegraphics[width=\textwidth]{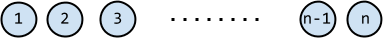}
\end{minipage}
\begin{minipage}[]{0.45\linewidth}
\hspace*{0.3in} $\Pr[\text{broadcaster} = 1] = \alpha$
\end{minipage}
\\
\begin{minipage}[]{1\linewidth}
\centering \vspace{.15in} (a) Random broadcasters model \vspace{.15in}
\end{minipage} 
\\
\begin{minipage}[]{0.45\linewidth}
\centering
\includegraphics[width=\textwidth]{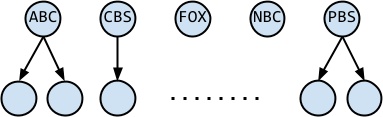}
\end{minipage}
\begin{minipage}[]{0.40\linewidth}
\hspace*{0.3in} $\Pr[\text{affiliate} = 1] = \alpha$ \\\\
\hspace*{0.3in} $\Pr[\text{broadcaster} = 1 \mid \text{affiliate} = 0] = 0$ \\
\hspace*{0.3in} $\Pr[\text{broadcaster} = 1 \mid \text{affiliate} = 1] = 1$ 
\end{minipage}
\\
\begin{minipage}[]{1\linewidth}
\centering \vspace{.15in} (b) Random affiliates model \vspace{.15in}
\end{minipage} 
\\
\begin{minipage}[]{0.45\linewidth}
\centering
\includegraphics[width=\textwidth]{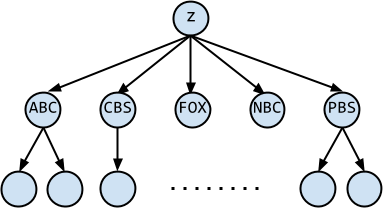}
\end{minipage}
\begin{minipage}[]{0.45\linewidth}
\hspace*{0.3in} $\Pr[z = 1] = 0.9$ \\\\
\hspace*{0.3in} $\Pr[\text{affiliate} = 1 \mid z = 0] = 0$ \\
\hspace*{0.3in} $\Pr[\text{affiliate} = 1 \mid z= 1] = \frac{\alpha}{0.9}$ \\\\
\hspace*{0.3in} $\Pr[\text{broadcaster} = 1 \mid \text{affiliate} = 0] = 0$ \\
\hspace*{0.3in} $\Pr[\text{broadcaster} = 1 \mid \text{affiliate} = 1] = 1$ 
\end{minipage}
\\
\begin{minipage}[]{1\linewidth}
\centering \vspace{.15in} (c) Correlated affiliates model
\end{minipage} 
\\
\caption{\it Bayes nets representing three classes of broadcaster participation models. In each model, the marginal probability of broadcaster non-participation is $\alpha$.}
\label{fig:models}
\end{figure}

All of the models above belong to the class of \emph{graphical models} (more specifically, the
class of Bayesian networks),
which are probabilistic models
that can be represented by a \emph{graph}, i.e. a set of nodes and edges. Each node in the graph represents a random variable,
and a directed edge from node $i$ to node $j$ denotes that the setting of variable $i$ depends on the setting
of variable $j$. For instance, in the random affiliates model, each affiliate group and broadcaster is represented by a node. There is a directed edge pointing from each affiliate group to each of its members.
These edges reflect the fact that the participation decision of each member is not made in isolation, but 
instead depends on some top-level event or influence corresponding to its affiliation. The graphical models of
the participation models we consider are illustrated in Figure \ref{fig:models}. 


The broadcaster revenue model is somewhat different, in the sense that the marginal probability of participation is no longer fixed at $\alpha$. This model is defined by two parameters, $\beta$ and $\gamma$. The parameter $\beta$ controls the fraction of broadcasters whose non-participation probability is above $\frac{1}{2}$, and the parameter $\gamma$ controls the amount by which we amplify the non-participation probability of network affiliates. Formally, the model is created as follows. Let $r$ denote the sorted vector of broadcaster revenues. Let $k$ denote the index that lies a $1 - \beta$ fraction of the way through of $r$. So a $\beta$ fraction of $r$ comes after $k$. We now subtract $r(k)$ from all elements of $r$. This has the effect of shifting the revenues so that a $\beta$ fraction will be above 0. We then divide all revenues by a scaling factor $\max_i |r(i)|/4$, so that all values lie in the range $[-4, 4]$. Next, we perform a sigmoidal transformation, and set $r(i)$ equal to $1/(1 + \exp(-r(i)))$. Finally, for all affiliates, we update $r(i) = r(i) \cdot \gamma$.

\subsection{Clique Algorithm}
\label{sec:clique-algo}

Algorithm \ref{alg:blockingset} provides a formal description of the alternate clique-based approach for determining feasibility that was outlined in Section \ref{sec:blockingset}.
\begin{algorithm}
\caption{\it Clique Algorithm}
\label{alg:blockingset}
\begin{algorithmic}
\State $n = 1672$ \Comment{Number of broadcasters}
\State $m = 38$ \Comment{Number of channels}
\State $c = m - \lfloor M/6 \rfloor$ \Comment{Number of available channels}
\State $S = \{s \in S \mid |s| \geq c+1\}$ \Comment{Set of blocking cliques}
\State $\text{infeasibilities} = \text{zcount} = 0$
\For {$i=1$ to $T$}
\State $x \leftarrow P(\vec{x})$ \Comment{Sample $x$ from $P(\vec{x})$}
\State $Z = \emptyset$
\For {$s \in S$}
\State $B = \{i \in S \mid x(i) = 1\}$
\If {$|B| \geq c+1$} \Comment{$B$ is a blocking set}
\State $Z \gets Z \cup B$ \Comment{Take union of blocking cliques}
\EndIf
\EndFor
\If {$|Z| > 0$} \Comment{Infeasibility occurred}
\State $\text{infeasibilities} \gets \text{infeasibilities} + 1$
\State $z \gets \text{zcount} + |Z|$
\EndIf
\EndFor
\State $p = 1 - \text{infeasibilities}/T$ \Comment{$p$ is probability of success}
\State $z = \text{zcount}/\text{infeasibilities}$ \Comment{$z$ is average size of blocking set union}
\end{algorithmic}
\end{algorithm} 
\end{document}